\documentclass[conference,anonymous]{IEEEtran}
\IEEEoverridecommandlockouts
\usepackage{cite}
\usepackage{amsmath,amssymb,amsfonts}
\usepackage{algorithmic}
\usepackage{graphicx}
\usepackage{textcomp}
\usepackage{xcolor}

\usepackage{booktabs} 
\usepackage{graphicx} 
\usepackage{array}    
\usepackage{tabularx}  
\usepackage{amsmath}  
\usepackage{color}
\usepackage{soul}
\usepackage{comment}
\usepackage{multirow}
\usepackage[breaklinks=true]{hyperref}

\def\BibTeX{{\rm B\kern-.05em{\sc i\kern-.025em b}\kern-.08em
    T\kern-.1667em\lower.7ex\hbox{E}\kern-.125emX}}
\begin{document}

\title{Patient Journey Ontology: Representing Medical Encounters for Enhanced Patient-Centric Applications}

\author{
    \IEEEauthorblockN{
        Hassan S. Al Khatib\textsuperscript{1}, 
        Subash Neupane\textsuperscript{1}, 
        Sudip Mittal\textsuperscript{1}, 
        Shahram Rahimi\textsuperscript{2},\\
        Nina Marhamati\textsuperscript{3},
        Sean Bozorgzad\textsuperscript{3}
    }
    \IEEEauthorblockA{
        \textsuperscript{1}Mississippi State University, Starkville, MS, USA \\
        \textsuperscript{2}University of Alabama, Tuscaloosa, AL, USA \\
        \textsuperscript{3}Potentia Analytics Inc, IL, USA \\
        Emails: \{hsa78, sn922\}@msstate.edu, mittal@cse.msstate.edu, srahimi1@ua.edu, \\ \{nina, sean\}@potentiaco.com
    }
}

\maketitle

\begin{abstract}
The healthcare industry is moving towards a patient-centric paradigm that requires advanced methods for managing and representing patient data. This paper presents a Patient Journey Ontology (PJO), a framework that aims to capture the entirety of a patient's healthcare encounters. Utilizing ontologies, the PJO integrates different patient data sources like medical histories, diagnoses, treatment pathways, and outcomes; it enables semantic interoperability and enhances clinical reasoning. By capturing temporal, sequential, and causal relationships between medical encounters, the PJO supports predictive analytics, enabling earlier interventions and optimized treatment plans. The ontology's structure, including its main classes, subclasses, properties, and relationships, as detailed in the paper, demonstrates its ability to provide a holistic view of patient care. Quantitative and qualitative evaluations by Subject Matter Experts (SMEs) demonstrate strong capabilities in patient history retrieval, symptom tracking, and provider interaction representation, while identifying opportunities for enhanced diagnosis-symptom linking. These evaluations reveal the PJO's reliability and practical applicability, demonstrating its potential to enhance patient outcomes and healthcare efficiency. This work contributes to the ongoing efforts of knowledge representation in healthcare, offering a reliable tool for personalized medicine, patient journey analysis and advancing the capabilities of Generative AI in healthcare applications.
\end{abstract}


\section{Introduction}
 
The transformation of healthcare towards a patient-centric model indicates the need for advanced methods to manage and represent patients' data. A crucial concept in this context is the ``patient journey,'' a term coined by the European Organization for Rare Diseases (EURORDIS)\cite{eurordis}, which refers to a patient's experience throughout an episode of care, from initial admission, diagnosis, and treatment to hospital discharge. For example, the journey for a cancer patient begins with an initial medical encounter where the patient consults a primary physician, who reviews the medical and social history, checks for symptoms, and runs some tests. The patient is then referred to a specialist for further diagnostic tests and treatments, including medical encounters with oncologists for chemotherapy and radiation therapy. This journey continues with regular follow-up visits to monitor recovery and manage post-treatment care. However, the current state of healthcare data is often siloed and unstructured within Electronic Medical Records (EMRs). This fragmentation makes it challenging to integrate data effectively, hindering the ability to gain a holistic view of a patient’s journey. Furthermore, without a structured semantic framework, it becomes challenging to piece together comprehensive insights from disparate data sources.

To address these challenges, this paper presents the \textbf{Patient Journey Ontology (PJO)}, a formal knowledge representation that systematically captures and structures the relevant entities (such as medical encounters, treatments, outcomes, etc.) and relationships within a patient's medical encounters. The ontology defines these entities and their relationships in a manner that supports comprehensive data integration and analysis \cite{guarino}, enabling insights crucial for optimizing patient care.

Ontologies are becoming increasingly popular as powerful tools for structuring knowledge in healthcare settings.  
Known examples include the Systematized Nomenclature of Medicine-Clinical Terms (SNOMED CT), the Gene Ontology, and the Unified Medical Language System (UMLS), each of which facilitates the integration and utilization of complex biomedical data\cite{bodenreider}. These ontologies enable the aggregation of diverse data elements such as medical histories, diagnostic information, treatment pathways, and patient outcomes, thus fostering a holistic understanding of patient care. Unlike FHIR \cite{ayaz}, which is primarily focused on the standardization of data exchange between healthcare systems, and OMOP-CDM \cite{biedermann}, which structures clinical data for large-scale observational research, the PJO provides a more comprehensive, semantically rich framework that supports in-depth knowledge representation and reasoning across a wide range of healthcare scenarios.

\begin{table*}[t]
\caption{Comparison of PJO with Existing Healthcare Standards}
\label{table:standards-comparison}
\centering
{\renewcommand{\arraystretch}{1.30}
\begin{tabular}{p{2.5cm}|p{3cm}|p{3cm}|p{3cm}|p{3cm}}
\hline
\textbf{Feature} & \textbf{PJO} & \textbf{FHIR} & \textbf{OMOP-CDM} & \textbf{SNOMED CT} \\
\hline
Primary Focus & Patient journey tracking & Data exchange & Observational research & Clinical terminology \\
\hline
Temporal Support & Extensive & Basic & Limited & None \\
\hline
Causal Relationships & Explicit & Implicit & Not supported & Limited \\
\hline
Care Coordination & Primary feature & Partial support & Not supported & Not applicable \\
\hline
Implementation & Ontology-based & REST API & Relational database & Terminology service \\
\hline
Use Cases & Journey analysis & Interoperability & Research analytics & Clinical coding \\
\hline
Extensibility & Highly extensible & Profiles \& extensions & Custom tables & Post-coordination \\
\hline
Integration Capacity & Standards-compatible & Native support & ETL required & Reference sets \\ \hline
\multicolumn{5}{c}{\textbf{PJO Alignment}} \\ \hline
\multicolumn{2}{c|}{UMLS \checkmark } & FHIR \checkmark  & SNOMED CT \checkmark  & ICD \checkmark  \\

\hline
\end{tabular}}
\end{table*}

Unlike FHIR and OMOP-CDM, which focus primarily on data exchange and observational research, respectively, the PJO specifically addresses the temporal and semantic representation of complete patient care journeys. The PJO extends beyond traditional ontologies through built-in support for care transitions, explicit causal relationships between medical events, and comprehensive healthcare provider relationship mapping. In addition, the PJO aligns with the UMLS by mapping its concepts to UMLS Concept Unique Identifiers (CUIs), leveraging standardized medical terminologies such as SNOMED CT and ICD. This mapping ensures semantic interoperability with existing healthcare frameworks and facilitates seamless integration with EMRs. Its semantic structure is optimized for Artificial Intelligence (AI) and Machine Learning (ML) applications through structured data representations that enable pattern recognition and predictive modeling, making it particularly valuable for clinical decision support systems and personalized care journey development. Table~\ref{table:standards-comparison} provides a detailed comparison of PJO with existing healthcare standards.

Despite the major developments in healthcare ontologies, there is a notable absence of ontologies specifically designed to describe a patient's journey. The lack of a dedicated PJO hinders the development of expert clinical AI systems, such as question-answering applications capable of providing answers based on a patient's unique context. It impedes the development of AI systems capable of delivering accurate, context-aware healthcare recommendations.

The paper has three main contributions:
\begin{itemize}
    \item It presents a new ontology designed explicitly to represent a patient journey in a structured and semantically rich manner.
    \item It details the patient journey components: medical encounters, intake forms, and multiple medical encounter tracking. Medical encounters document patient-provider interactions, including symptoms, diagnoses, treatments, and outcomes. Patient intake forms collect essential personal and medical information for personalized care. Through multiple medical encounters, the system shows how ongoing interactions shape clinical decisions and patient outcomes, illustrating the complete healthcare experience.
    \item It provides a detailed evaluation of the PJO, highlighting its efficacy and potential impact on healthcare outcomes and efficiency.
\end{itemize}

This paper is structured as follows: Section \ref{background} reviews related work on artificial intelligence, ontologies, and their applications in healthcare. Section \ref{patient_journey} details our patient profile, presents the detailed design of the PJO, including its main classes, subclasses, properties, and relationships, and includes instances to show how the patient data fits into the PJO. Section \ref{ontology_evaluation} discusses the evaluation methodologies for the ontology’s efficacy and applicability. Finally, Section \ref{conclusion} concludes with a summary of findings and directions for future research.

\begin{figure*}[h]
  \centering
  \includegraphics[width=0.95\linewidth]{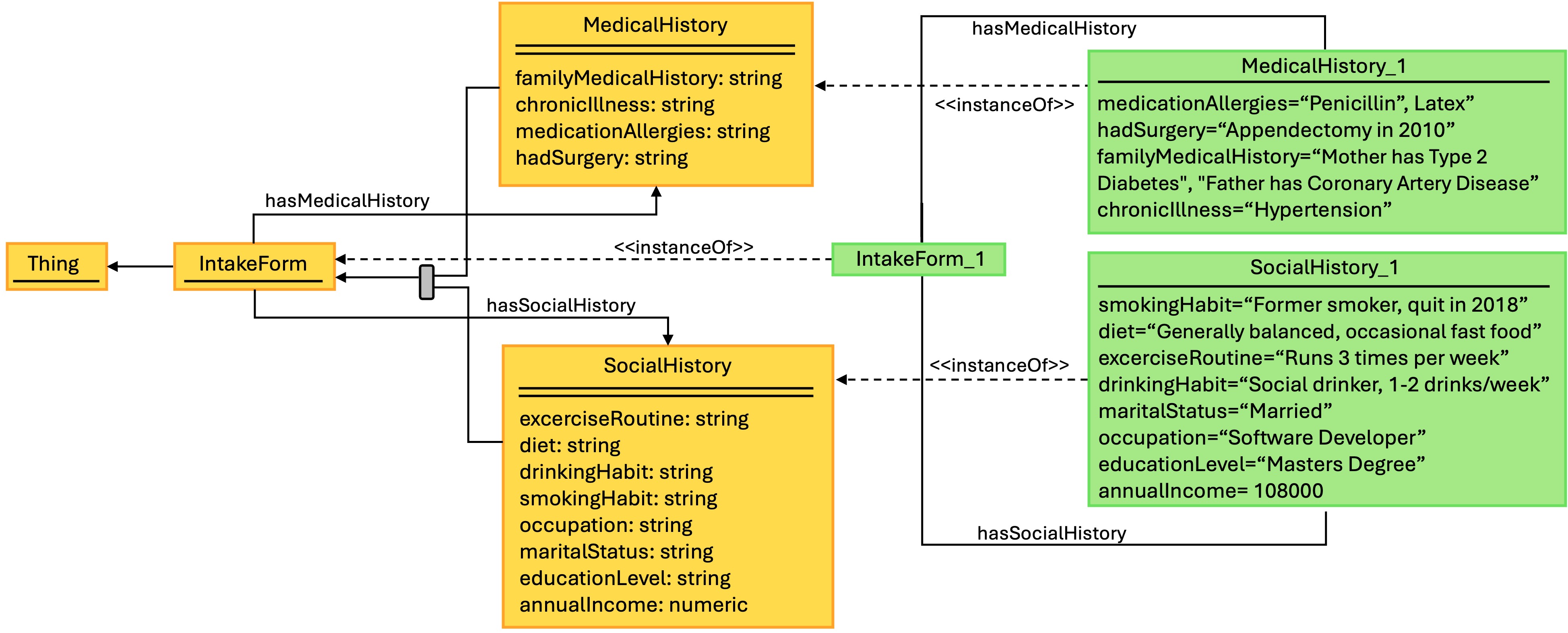}
  \caption{Diagram of the Intake Form Ontology with instances illustrating its components and a specific instance that captures a patient’s comprehensive social and medical history. This example presents John Doe, a patient with a past surgery and a current chronic condition, as well as lifestyle factors such as smoking and drinking habits, diet, and exercise routines.}
  \label{fig:intake}
\end{figure*}

\section{Background}
\label{background}
This section summarizes key applications of AI and ontologies in healthcare, focusing on their role in patient journey representation and analysis.

\subsection{AI in Healthcare}

AI technologies, particularly Machine Learning (ML) and Natural Language Processing (NLP), are transforming healthcare by improving patient care delivery and operational efficiency. Mishra et al. \cite{mishra} demonstrated how AI enhances patient-provider interactions through real-time data analytics and clinical decision support. As discussed by the authors in \cite{stanfill}, AI significantly impacts health information management by automating tasks, providing real-time insights, and predicting outcomes, thereby increasing efficiency and improving patient experience. 

Recent applications show the particular value of AI in personalized medicine. Abatal \& Korchi \cite{abatal} illustrate how AI enables personalized treatment strategies by integrating genomic data with medical records and outcome measures. Similarly, Kim et al. \cite{ykim} demonstrate the effectiveness of AI in specialized domains such as thoracic radiology. These applications highlight the potential of AI to support the analysis and management of the complete patient journey.

\subsection{Ontologies in Healthcare}

Ontologies in healthcare serve as formal representations of concepts, entities, and their interrelationships, essential for structuring and organizing knowledge within the domain. They enable semantic interoperability and facilitate data integration, ensuring consistency in data interpretation and improving information retrieval processes \cite{ocan}, \cite{pckg}, \cite{onteng}. Biomedical ontologies encode various aspects, including phenotypes, diseases, and medical terminologies, thus providing a standard framework for data storage and exchange \cite{slater, adel}. Different ontologies have been created to meet specific demands, such as dealing with chronic diseases such as diabetes or integrating social networks into semantic technologies to create personalized models for patients who suffer from multiple chronic diseases \cite{kaldoudi, tarabi}.

Dynamic ontologies are essential in reflecting the everevolving nature of medical knowledge and practice. They allow integration and interoperability among various healthcare data types to accommodate real-time updates that follow the latest clinical guidelines, research findings, and patient care protocols. These complexities can be better handled by combining Service-Oriented Architecture (SOA) with dynamic ontologies in healthcare environments, particularly during emergencies \cite{zeshan}. These systems enable seamless data exchange, integration, and reuse, enhancing significantly the efficiency and flexibility of healthcare delivery.

\begin{figure*}[h]
  \centering
  \includegraphics[width=0.95\linewidth]{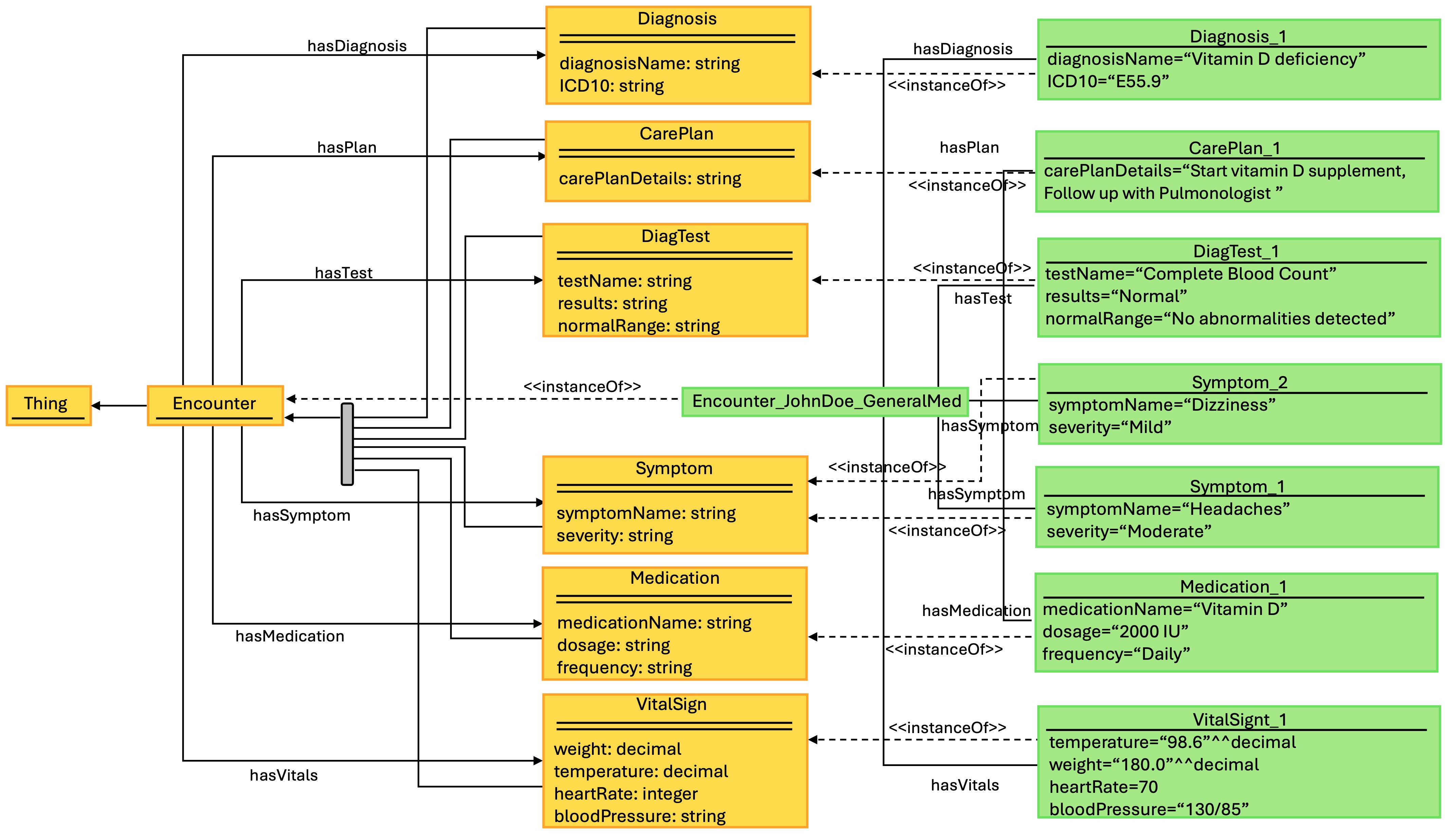}
  \caption{Diagram of the Medical Encounter Ontology with instances illustrating the comprehensive components and their interrelations. This diagram includes an instance demonstrating John Doe's first medical encounter in general medicine, specifically detailing the process for diagnosing and managing Vitamin D deficiency. The ontology captures various elements such as symptoms, vital signs, diagnostic tests, diagnosis, prescribed medication, and recommended follow-up care, providing a structured framework that effectively represents complex medical data.}
  \label{fig:encounter}
\end{figure*}

\section{Patient Journey}
\label{patient_journey}
The patient journey encompasses all medical encounters from birth to death \cite{trebble}, including regular visits, hospital admissions, treatments, and ongoing care. These touchpoints create a continuous timeline that supports immediate patient management and a greater understanding of health trends and treatment outcomes.

In this section, we discuss the different components of the patient journey, such as ``Medical Encounters'', each unique interaction with healthcare providers described individually; ``Intake Form'', where personal history is recorded along with relevant details about the current condition(s) being treated or diagnosed; and the concept of ``Multiple Encounters,'' illustrating how ongoing medical encounters influence decision making and outcomes. Each element contributes to a holistic view of patient care, informing better health practices and policies.

\subsection{Patient Data}
In this study, we use anonymized and deidentified patient data to comply with the Health Insurance Portability and Accountability Act (HIPAA) regulations \cite{hippa}. For example, we created a comprehensive profile for a US-based patient, John Doe, a 39-year-old white male with a generally healthy profile and mild chronic disease. This profile was developed with the insight of an experienced physician to accurately reflect the complexity of real medical records and ensure that it is representative of the general population. It includes an overview of John's medical conditions and his proactive health management.
His intake form shows that our patient was diagnosed with hypertension, which he manages with medication, and he also deals with seasonal allergies. His family medical history includes chronic diseases such as type 2 diabetes and coronary artery disease. Socially, John's lifestyle as a former smoker, moderate alcohol consumption, and regular physical activities like jogging reflect his commitment to health.

John's journey consisted of four medical encounters, documented through detailed narratives of his medical consultations. These include his first medical encounter for general medicine consultation for symptoms such as persistent headaches, followed by a second medical encounter for specialized visits for conditions such as mild sleep apnea, diagnosed following a sleep study. His third and fourth medical encounters with allergists also highlight his struggle with and management of seasonal allergies. The data from these medical encounters will be showcased as instances in this section of the paper, illustrating the dynamic and interconnected aspects of his journey. The primary objective of this study is to comprehensively represent John Doe's medical journey using the PJO.

\begin{figure*}[htb]
  \centering
  \includegraphics[width=0.95\linewidth]{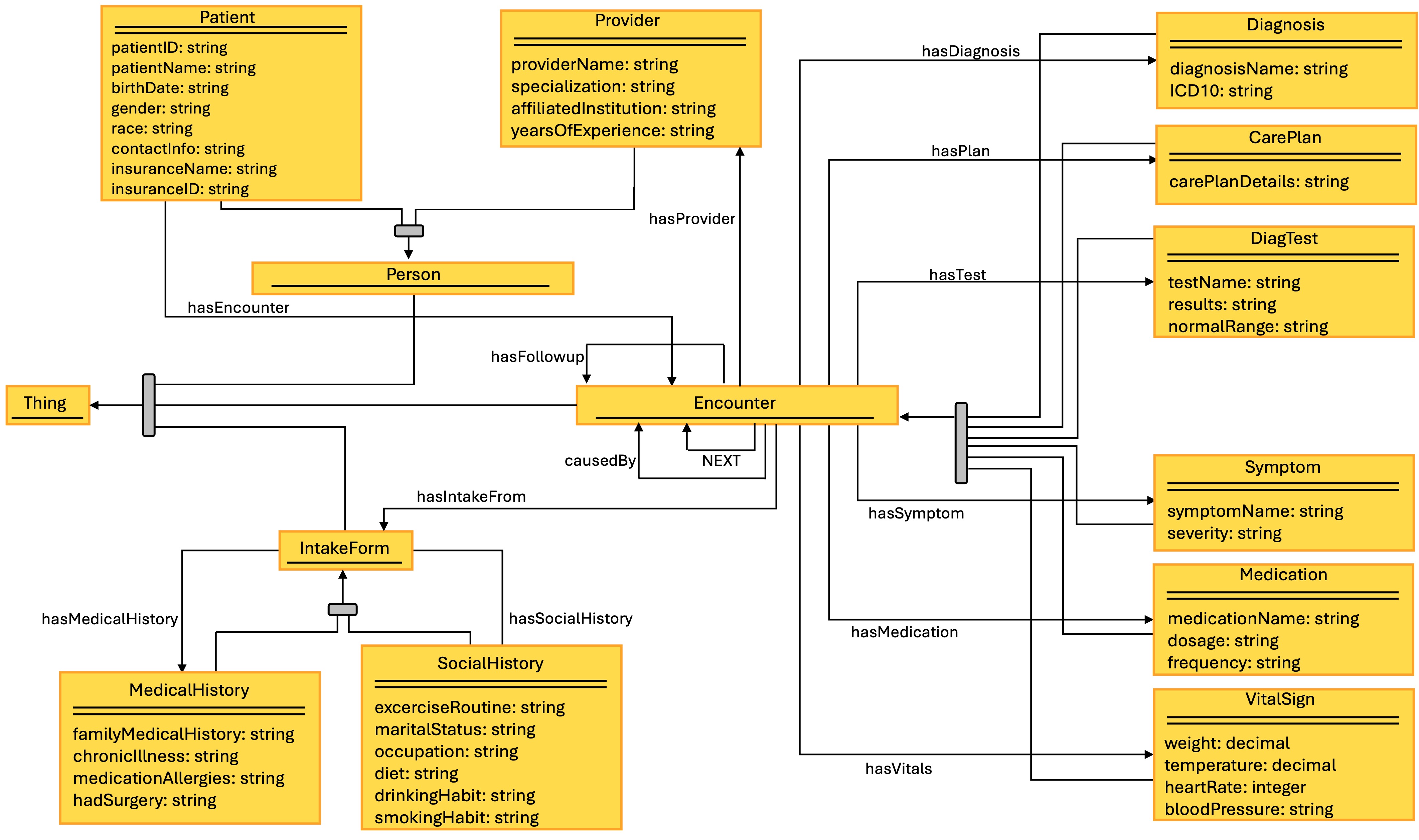}
  \caption{Diagram of the developed PJO, outlining its comprehensive structure, including all classes, subclasses, properties, and temporal and causal relationships that map a patient's healthcare journey. The ontology encapsulates elements from initial patient intake through multiple medical encounters, covering medical history, symptoms, diagnostics, treatments, and follow-ups to provide a systematic framework for capturing and analyzing patient care processes.}
  \label{fig:journey}
\end{figure*}

\begin{figure*}[h]
  \centering
  \includegraphics[width=0.95\linewidth]{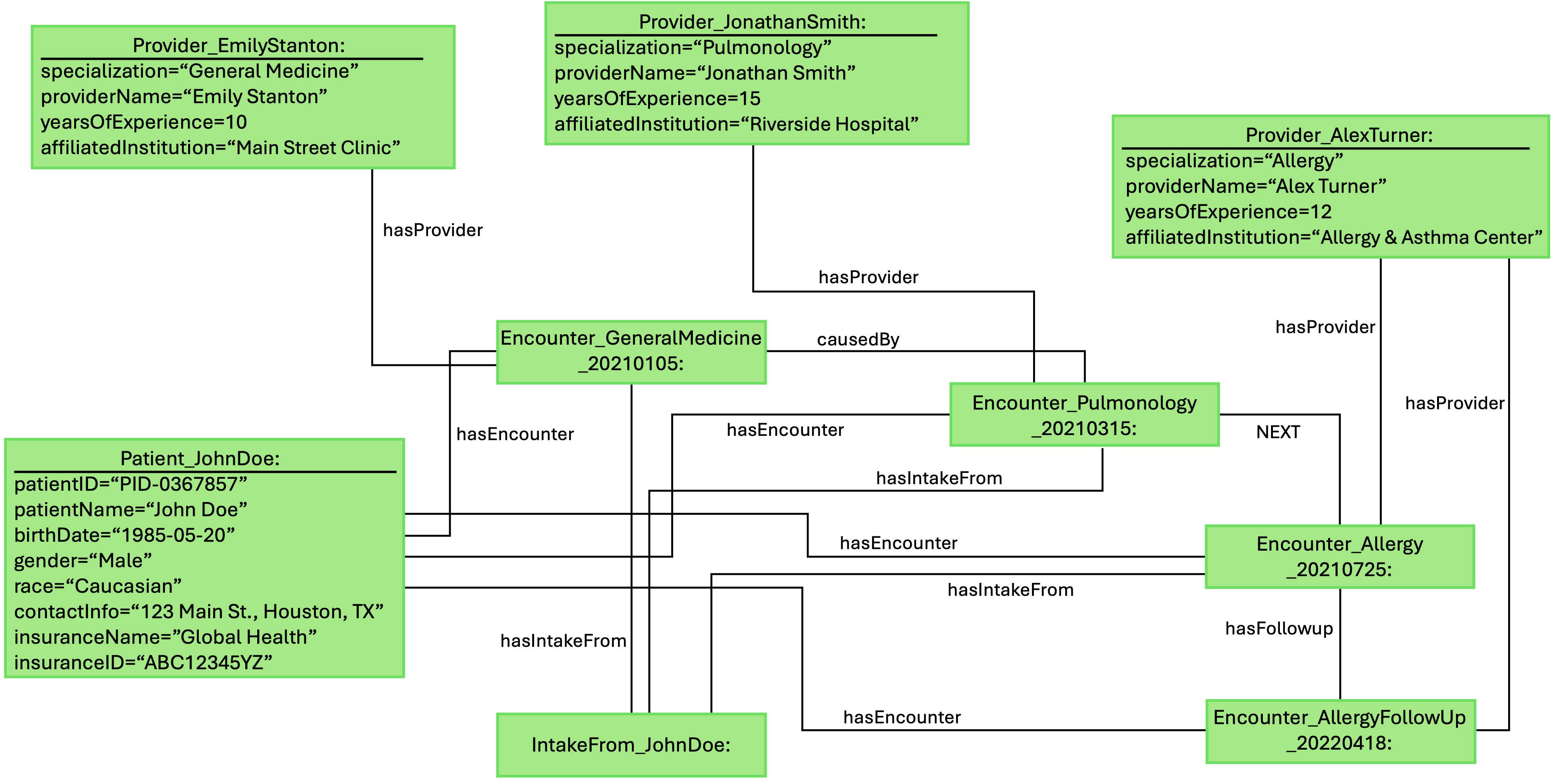}
  \caption{Diagram of PJO instances demonstrating a sequence of four medical encounters in John Doe's journey. The figure illustrates the temporal, causal, and sequential relationships among these medical encounters, from the first general medicine consultation through follow-up allergy consultations, highlighting the PJO's capability to trace condition progression and link comprehensive patient data.}
  \label{fig:instance}
\end{figure*}

\subsection{Patient Intake Form}

Outpatient clinics use a patient intake form during the initial medical encounter to gather personal and medical information essential for individualized care. It includes identification details like name, date of birth, and contact information, along with medical history covering past illnesses, ongoing conditions, medications, allergies, and family disease background. Lifestyle choices, such as smoking, alcohol consumption, diet, and exercise habits, are also recorded, offering insight into overall wellness and the reason for the visit. A well-designed intake form supports healthcare providers in making informed decisions for both immediate and long-term treatment, enhancing patient safety, satisfaction, and the efficiency of care.

\subsubsection{Ontology Structure for Patient Intake Form} 
The proposed patient intake form ontology provides an effective representation for patient data. It encapsulates patient information in a structured framework comprising main classes, subclasses, and properties representing the form's diverse elements.

\begin{itemize}
    \item \textbf{Main Class:} The \small{\texttt{IntakeForm}} class serves as a primary class, representing the patient intake form. It encapsulates medical and social history information, providing a unified view of the patient's data.
    
    \item \textbf{Subclasses of IntakeForm:} The IntakeForm contains two primary subclasses that organize patient information. The first subclass, \texttt{IntakeForm:MedicalHistory} (UMLS CUI: C0262926), encompasses detailed information about the patient's health background. This includes past medical events, chronic conditions, medications, allergies, and family medical history. For healthcare providers, this subclass is essential in understanding the patient's medical background and identifying potential health risks. The MedicalHistory subclass implements several key properties including \small{\texttt{hadSurgery}} for recording surgical history, \small{\texttt{chronicIllness}} for ongoing conditions, \small{\texttt{medicationAllergies}} for drug sensitivities, and \small{\texttt{familyMedicalHistory}} for hereditary health factors.
    
    The second subclass, \small{\texttt{IntakeForm:SocialHistory}} (UMLS CUI: C3714536), focuses on capturing the patient's lifestyle and social determinants of health. This subclass documents various aspects of the patient's daily life and social circumstances that may impact their health outcomes. It includes properties such as \small{\texttt{smokingHabit}} and \small{\texttt{drinkingHabit}} for tracking substance use, \texttt{diet} for dietary patterns, and \small{\texttt{exerciseRoutine}} for physical activity levels. Additionally, it captures socioeconomic factors through properties like \small{\texttt{maritalStatus}}, \texttt{occupation}, \small{\texttt{educationLevel}}, and \texttt{annualIncome}, providing a comprehensive view of the patient's social context.
    
    \item \textbf{Relationships:} To define the interconnections between components, we establish specific relationships within the ontology. The \small{\texttt{hasMedicalHistory}} relationship links an intake form to its corresponding medical history instances, while the \small{\texttt{hasSocialHistory}} relationship connects an intake form to its associated social history instances. These relationships ensure that the intake form captures both medical and social aspects of a patient's data comprehensively.
\end{itemize}

\subsubsection{Instance Creation} 
To illustrate the application of the \small{\texttt{IntakeForm}} ontology, we created example instances for \small{\texttt{MedicalHistory}} and \small{\texttt{SocialHistory}}. As depicted in Figure 1, the patient's \texttt{IntakeForm-1} includes a \small{\texttt{MedicalHistory-1}} instance detailing a previous appendectomy, hypertension, allergies to penicillin and latex, and a family history of diabetes and heart disease. In contrast, the patient's \small{\texttt{SocialHistory-1}} instance captures their status as a former smoker, a social drinker with a balanced diet, and a regular exercise routine. This example demonstrates how the ontology can be populated with real-world data, comprehensively and systematically capturing all necessary information from a patient's intake form to facilitate efficient and individualized healthcare.

\subsection{Medical Encounter}
In the healthcare context, a medical encounter represents an event where a patient interacts with healthcare providers, typically involving visits to doctors, hospitals, or other treatment facilities. This project focuses on outpatient visits, treating each medical encounter as a distinct event where medical processes are documented and actions are taken. We define a medical encounter as a specific instance of patient care initiated when a patient visits a healthcare setting for examination, treatment, or consultation. These interactions are crucial for diagnosing conditions, planning treatments, and dynamically managing patient health.

\subsubsection{Ontology Structure for Medical Encounter} 
To represent the comprehensive data for a medical encounter, we developed an ontology that encapsulates various aspects of a medical encounter, providing a structured framework for data storage and retrieval, with main classes, subclasses, properties, and relationships representing its diverse elements.

\begin{itemize}
    \item \textbf{Main Class:} The \small{\texttt{Encounter}} (UMLS CUI: C3714536 represented as \emph{Patient Visit} and Encounter in FHIR) class serves as the primary class in the ontology, representing the medical encounter as a whole. It encompasses various components such as symptoms, vital signs, diagnoses, medications, care plans, and diagnostic tests, providing a comprehensive view of the patient's interaction with healthcare providers.
    
    \item \textbf{Subclasses of Encounter:} The main class comprises seven subclasses that together capture the complete medical encounter. The \small{\texttt{Encounter:Diagnosis}} (UMLS CUI: C0011900) subclass includes information about the medical diagnosis given during the medical encounter, capturing properties such as \small{\texttt{diagnosisName}} and the \small{\texttt{ICD10}} code associated with the diagnosis. For symptom documentation, the \small{\texttt{Encounter:Symptom}} (UMLS CUI: C3540840 represented as Sign or Symptom) subclass records the symptoms reported by the patient during the encounter, including the \small{\texttt{symptomName}} and \small{\texttt{severity}} of each symptom. The \small{\texttt{Encounter:Medication}} subclass details any medications prescribed or recommended during the encounter, tracking the \small{\texttt{medicationName}} (UMLS CUI: C5939153) , \small{\texttt{dosage}}, and \small{\texttt{frequency}}. To monitor patient health indicators, the \small{\texttt{Encounter:VitalSign}} (UMLS CUI: C0518766) subclass captures vital signs recorded during the encounter, including \small{\texttt{bodyTemperature}}, \small{\texttt{bloodPressure}}, \small{\texttt{weight}}, and \small{\texttt{heartRate}}. The \small{\texttt{Encounter:DiagTest}} (UMLS CUI: C0086143) subclass maintains information about diagnostic tests performed during the encounter, capturing the \small{\texttt{testName}}, \small{\texttt{results}}, and \small{\texttt{normalRange}}. Finally, the \small{\texttt{Encounter:CarePlan}} UMLS CUI: C2735110 (Care process or plan) subclass details the encounter's recommendations, including lifestyle changes, therapy, or follow-up actions, providing a comprehensive plan for ongoing patient care.
    
    \item \textbf{Relationships:} To define how the components of a medical encounter are interconnected, we establish several relationships within the ontology. The \small{\texttt{hasSymptom}} relationship links an \small{\texttt{Encounter}} to its associated \small{\texttt{Symptom}} instances, while \small{\texttt{hasDiagnosis}} connects an \small{\texttt{Encounter}} to its corresponding \small{\texttt{Diagnosis}} instances. The \small{\texttt{hasMedication}} relationship associates an \small{\texttt{Encounter}} with its \small{\texttt{Medication}} instances, and \small{\texttt{hasVitals}} links to relevant \small{\texttt{VitalSign}} instances. Additionally, the \small{\texttt{hasTest}} relationship connects to related \small{\texttt{DiagTest}} instances, and \small{\texttt{hasPlan}} links to corresponding \small{\texttt{CarePlan}} instances. These relationships collectively ensure that all aspects of a medical encounter are properly connected and traceable within the ontology structure.
    
\end{itemize}

\subsubsection{Instance Creation} 
To demonstrate the application of the \small{\texttt{Encounter}} ontology, we created example instances for each component of a medical encounter. As presented in Figure 2, a patient underwent ``Encounter-1" where the patient exhibited several symptoms: (Symptom-1 = ``Headaches") and (Symptom-2 = ``Dizziness"). Patient's vital signs, ``VitalSign-1", were recorded, and diagnostic tests, specifically a (DiagTest-1 = ``Complete Blood Count-CBC"), were conducted and had normal results. The physician's evaluation (Diagnosis-1) identified a Vitamin D deficiency. Consequently, the patient was prescribed a Vitamin D supplement (2000IU) daily (Medication-1). Additionally, a follow-up with a Pulmonologist was recommended as part of the care plan (CarePlan-1). These instances illustrate the practical utility of ontology in capturing structured and comprehensive data reflective of real-world medical encounters.

\subsection{Patient's Journey - Multiple Medical Encounters}
The patient journey is event-centric, involving medical encounters, decisions, and experiences for health management represented through an ontology. This ontology provides a holistic view of a patient's interactions with the healthcare system, from initial consultations to ongoing treatments, illustrating how different medical encounters influence the journey.

\subsubsection{Patient Journey Ontology}
\paragraph{Main Classes} The PJO consists of three primary classes. In the previous sections, we described the classes \small{\texttt{Medical Encounter}} and \small{\texttt{IntakeForm}}, along with their associated subclasses. The third main class, \small{\texttt{Person}}, is a fundamental element that categorizes the individuals involved in healthcare care processes, with two key subclasses: \small{\texttt{Patient}} and \small{\texttt{Provider}}.

\paragraph{Subclasses} The ontology defines two key subclasses under the \small{\texttt{Person}} class: \small{\texttt{Person:Patient}} (UMLS CUI: C0030705) and \small{\texttt{Person:Provider}} (UMLS CUI: C2735026 represented as Primary care provider). The \small{\texttt{Person:Patient}} subclass represents individuals receiving medical treatment or consultation, characterized by properties such as \small{\texttt{patientID}}, \small{\texttt{patientName}}, \small{\texttt{birthDate}}, \small{\texttt{race}}, \small{\texttt{gender}}, and \small{\texttt{contactInformation}}, which includes details like address, phone number, email, and emergency contact information. Additionally, it captures health insurance (UMLS CUI: C0021682) details through properties such as \small{\texttt{insuranceName}} and \small{\texttt{insuranceID}}. On the other hand, the \small{\texttt{Person:Provider}} subclass refers to healthcare professionals responsible for delivering medical services. It includes properties such as \small{\texttt{providerName}}, \small{\texttt{specialization}}, \small{\texttt{affiliatedInstitution}}, and \small{\texttt{yearsOfExperience}}, providing a comprehensive representation of healthcare providers' credentials and roles.

\paragraph{Relationships} The ontology establishes several core relationships to define the interactions between key entities. The \small{\texttt{hasEncounter}} relationship links a \small{\texttt{Patient}} to their associated \small{\texttt{Medical Encounter}} instances, capturing the patient's individual healthcare interactions. The \small{\texttt{hasIntakeForm}} relationship connects a \small{\texttt{Patient}} to their corresponding \small{\texttt{IntakeForm}} instances, enabling the integration of comprehensive personal and medical history data. Additionally, the \small{\texttt{hasProvider}} relationship links a \small{\texttt{Medical Encounter}} to its associated \small{\texttt{Provider}} instances, representing the healthcare professionals involved in delivering medical services during specific medical encounters.

\paragraph{Temporal, Causal, and Sequential Relationships} To accurately represent the dynamics and progression of medical conditions and interventions within patient's medical encounters, we employ three types of relationships:

\begin{itemize}
    \item \textbf{Temporal Relationships:} The \small{\texttt{hasFollowup}} relationship represents the continuation or progression of a condition across multiple medical encounters, enabling the tracking of a patient's evolving condition through diagnoses, treatments, and follow-up visits.
    
    \item \textbf{Causal Relationships:} The \small{\texttt{causedBy}} relationship illustrates cause-and-effect dynamics, such as when a diagnosis or treatment decision leads to a referral, directly impacting the patient's treatment plan.
    
    \item \textbf{Sequential Relationships:} \small{\texttt{NEXT}} relationships capture the chronological sequence of medical encounters that are not directly causally related or classified as follow-ups, ensuring a linear and complete representation of the patient's medical history.
\end{itemize}

\subsubsection{Instance Creation} 
Figure 4 presents instances of the PJO capturing the different relationships between medical encounters in a patient's journey. For example, John Doe's journey, documented in the PJO, begins with a General Medicine consultation (with encounter\_id: ``Encounter-GeneralMedicine-20210105") for headaches and dizziness, leading to a diagnosis of vitamin D deficiency and a referral to a Pulmonologist, who (in encounter\_id: ``Encounter-Pulmonology-20210315") identifies mild sleep apnea. Subsequent allergy consultations (encounter\_id:``Encounter-Allergy-20210725" and encounter\_id: ``Encounter-AllergyFollowUp-20220418") address seasonal allergies and monitor the effectiveness of treatment.

The PJO captures the temporal, causal and sequential relationships between these medical encounters, allowing healthcare providers to trace the progression of the condition, understand the impacts of treatment, and make informed decisions. By linking intake form (``IntakeForm-JohnDoe") data with the different medical encounters, the PJO provides a holistic view of the patient's health, enhancing continuity of care, and improving outcomes.

\section{Ontology Evaluation}
\label{ontology_evaluation}
Evaluating an ontology for a patient journey requires rigorous assessment of its efficacy, accuracy, and applicability in real-world healthcare settings \cite{gangemi}. Our evaluation methodology combines quantitative metrics and qualitative expert assessment to ensure comprehensive coverage and clinical utility.

\subsection{Evaluation Methodology}

We validated the ontology's ability to accurately represent and query patient journey data across diverse clinical scenarios through \emph{competency questions} and \emph{scenario-based testing} \cite{brank, uschold, gruninger}. Five SMEs, all physicians with 11 to over 20 years of experience across pediatrics, cardiothoracic surgery, and infectious diseases, participated in our systematic evaluation. Using a comprehensive 20-question survey, they assessed multiple dimensions of the ontology, including its completeness, structural integrity, clinical accuracy, and domain representation. The evaluation also focused on practical aspects such as healthcare applicability, patient journey representation, and integration capabilities with existing systems.

\subsection{Quantitative Assessment}

Our statistical analysis yielded a Fleiss' Kappa value of 0.6117 (95\% CI: 0.454 to 0.770), indicating substantial agreement among the evaluators regarding the ontology's effectiveness \cite{mchugh}. The standard error of 0.1303 reflects the precision of our Kappa estimate, providing strong statistical support for the reliability of the expert assessments.

\begin{figure}[h]
  \centering
  \includegraphics[width=0.9\linewidth]{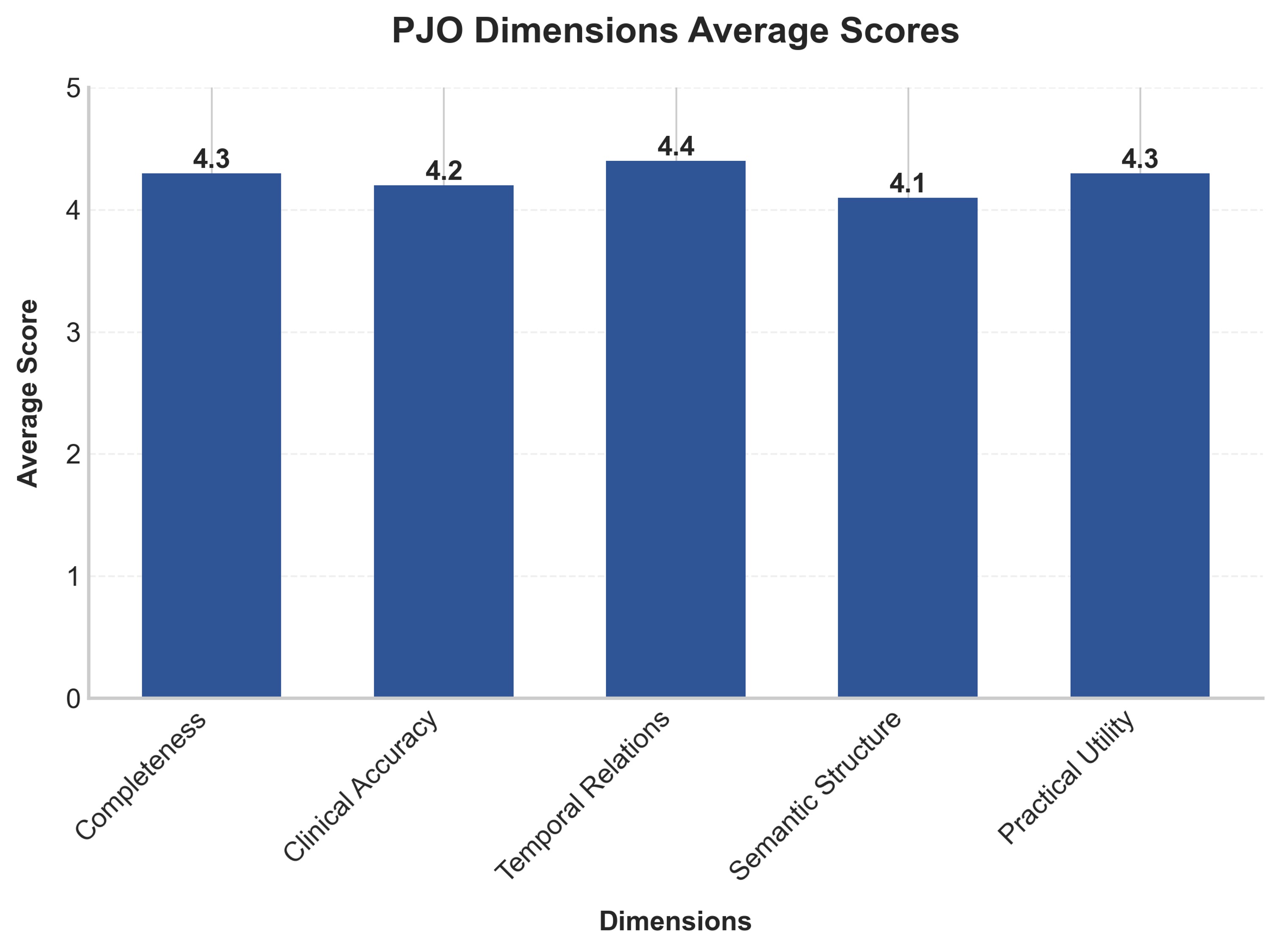}
  \caption{Average evaluation scores across PJO dimensions. The evaluation was conducted using a 5-point Likert scale, with higher scores indicating better performance. All dimensions scored above 4.0, with temporal relationships achieving the highest score (4.4) and Semantic Structure the lowest (4.1)}
  \label{fig:quantitative}
\end{figure}

Further statistical analysis of the SME evaluations revealed consistently high ratings across multiple ontology dimensions (Figure \ref{fig:quantitative}). The mean rating across all evaluation criteria was 4.26 out of 5.0, with relatively low variance across dimensions (SD = 0.12), indicating strong consensus among experts about the ontology's quality. The tight clustering of scores between 4.1 and 4.4 suggests high reliability in these positive assessments. Particularly noteworthy was the strong agreement among experts regarding the ontology's completeness and practical utility, with over 85\% of responses falling in the ``agree" or ``strongly agree" categories.

The evaluation also examined the ontology's ability to represent complex patient journeys, with experts assessing both structural integrity and clinical accuracy. The data showed particularly strong agreement regarding the ontology's capability to capture temporal relationships (mean rating 4.4) and its effectiveness in representing clinical decision points (mean rating 4.2), while the practical utility score of 4.3 demonstrates strong potential for real-world implementation.

\subsection{Qualitative Assessment}

The qualitative evaluation revealed important insights into the ontology's practical applicability and effectiveness. SME feedback consistently highlighted the PJO's comprehensive coverage of patient journey elements, with particular emphasis on its robust handling of temporal and causal relationships. Experts specifically noted the ontology's effectiveness in capturing the nuanced progression of patient care across multiple medical encounters, providing a cohesive framework for understanding complex patient journeys. For example, when evaluating competency questions like \emph{``Can the ontology track the progression of a patient's symptoms over time?"} and \emph{``Can the ontology link symptoms to possible diagnoses accurately?"}, SMEs indicated strong capabilities in symptom tracking but noted potential improvements needed in diagnosis-symptom linking relationships.

Through detailed analysis of expert feedback, several key strengths emerged. The ontology's ability to represent complex care transitions was particularly praised, with experts noting its effectiveness in maintaining semantic consistency across different types of medical encounters. The implementation of causal relationship mapping was identified as a distinctive advantage, enabling more sophisticated analysis of care patterns and outcomes. Experts specifically commented on the ontology's potential to enhance clinical decision support systems through its structured representation of patient journey data.

The evaluation also identified areas for future enhancement. Multiple reviewers suggested expanding the ontology's coverage of specialized medical domains, particularly in areas such as chronic disease management and emergency care journeys. A cardiothoracic surgeon among the evaluators noted that while the ontology effectively supports various scenarios, additional validation across diverse clinical contexts would further strengthen its applicability.

Our evaluation confirms that the PJO effectively represents patient journeys, backed by statistical validation and expert assessment. Expert feedback provides direction for future improvements, while combined quantitative and qualitative results validate the ontology's potential to enhance care coordination and healthcare delivery.

\section{Conclusion \& Future Work}
\label{conclusion}
Patient-centric care is fundamental in modern healthcare, ensuring treatments and interventions are tailored to individual needs. The proposed PJO advances this goal by integrating diverse patient data (such as medical histories, diagnoses, treatments, and outcomes) into a unified semantic representation, improving semantic interoperability and clinical reasoning. Through rigorous evaluation by SMEs, we have demonstrated that the PJO enables predictive analytics, early interventions, and optimized care plans, contributing significantly to the broader efforts in advancing knowledge representation within healthcare. The ontology's structure effectively captures the temporal, causal, and sequential relationships between medical events, providing healthcare providers with a comprehensive view of patient care trajectories and enabling more informed decision-making. SME evaluation of key competency questions confirmed strong capabilities in patient history retrieval, symptom tracking, and provider interaction representation, though opportunities exist to enhance diagnosis-symptom linking relationships and follow-up care recommendation generation.

While our research demonstrates the PJO's effectiveness, the study focused primarily on outpatient care using a single, US-based patient profile. Future work will address these limitations by expanding the dataset to include a broader patient population and incorporating global interoperability standards like the Global Patient Summary (GPS) and International Patient Summary (IPS). Additional improvements will include more detailed patient interactions, physical examinations, and symptom tracking capabilities. These enhancements will further improve the ontology's ability to support comprehensive patient journey analysis and care optimization, ultimately contributing to improved patient outcomes and more efficient healthcare systems.


\end{document}